\documentclass[aps,prx,reprint,longbibliography,nobalancelastpage,superscriptaddress]{revtex4-1}

\usepackage{amsmath,amssymb}
\usepackage{graphicx}
\usepackage{xcolor}
\usepackage{mathrsfs}
\usepackage{hyperref}
\usepackage{accents}
\usepackage{newtxtext}
\usepackage[slantedGreek,cmintegrals]{newtxmath}
\hypersetup{colorlinks,linkcolor={blue!80!black},citecolor={blue!80!black},urlcolor={blue!80!black}}

\renewcommand{\vec}[1]{\boldsymbol{#1}}

\renewcommand{\pi}{\uppi}

\DeclareMathAlphabet{\mathcal}{OMS}{cmsy}{m}{n}
\DeclareMathAlphabet{\mathcalbf}{OMS}{cmsy}{b}{n}
\newcommand{\mat}[1]{\mathsf{#1}}
\renewcommand{\leq}{\leqslant}
\renewcommand{\geq}{\geqslant}

\newcommand{\figref}[2]{[Fig.~\hyperref[#1]{\ref*{#1}(#2)}]}
\newcommand{\figrefi}[2]{[Fig.~\hyperref[#1]{\ref*{#1}(#2)}, inset]}
\newcommand{\textfigref}[2]{Fig.~\hyperref[#1]{\ref*{#1}(#2)}}
\newcommand{\wholefigref}[1]{(Fig.~\ref{#1})}
\newcommand{\wholefigrefi}[1]{(Fig.~\ref{#1}, inset)}
\newcommand{\textwholefigref}[1]{Fig.~\ref{#1}}

\begin{document}
\title{Shape-Shifting Polyhedral Droplets}
\author{Pierre A. Haas}
\affiliation{Department of Applied Mathematics and Theoretical Physics, Centre for Mathematical Sciences, \\ University of Cambridge, 
Wilberforce Road, Cambridge CB3 0WA, United Kingdom}
\author{Diana Cholakova}
\author{Nikolai Denkov}
\affiliation{Department of Chemical and Pharmaceutical Engineering, Faculty of Chemistry and Pharmacy, \\ University of Sofia, 1164 Sofia, Bulgaria}
\author{Raymond E. Goldstein}
\affiliation{Department of Applied Mathematics and Theoretical Physics, Centre for Mathematical Sciences, \\ University of Cambridge, 
Wilberforce Road, Cambridge CB3 0WA, United Kingdom}
\author{Stoyan K. Smoukov}
\affiliation{School of Engineering and Materials Science, Queen Mary, University of London, \\ Mile End Road, London E1 4NS, United Kingdom}
\date{\today}%
\begin{abstract}
Cooled oil emulsion droplets in aqueous surfactant solution have been observed to flatten into a remarkable host of polygonal shapes with straight edges and sharp corners, but different driving mechanisms --- (i) a partial phase transition of the liquid bulk oil into a plastic rotator phase near the droplet interface and (ii) buckling of the interfacially frozen surfactant monolayer enabled by drastic lowering of surface tension --- have been proposed. Here, combining experiment and theory, we analyse the hitherto unexplored initial stages of the evolution of these `shape-shifting' droplets, during which a polyhedral droplet flattens into a polygonal platelet under cooling and gravity. Using reflected-light microscopy, we reveal how icosahedral droplets evolve through an intermediate octahedral stage to flatten into hexagonal platelets. This behaviour is reproduced by a theoretical model of the phase transition mechanism, but the buckling mechanism can only reproduce the flattening if surface tension decreases by several orders of magnitude during cooling so that the flattening is driven by buoyancy. The analysis thus provides further evidence that the first mechanism underlies the `shape-shifting' phenomena.
\end{abstract}
\maketitle

\section{Introduction}
The culmination of the geometric preoccupations of Ancient Greece was doubtless the classification of the five platonic solids \cite{* [] [{, Book XIII; for a modern translation, see e.g. }] euclid, *euclid2}. It is topology, however, that dictates that one of their number, the icosahedron, should abound in nature, among the shapes of virus capsids and other biological structures~\cite{bowick13}: Euler's formula implies the formation of at least twelve topological defects in a hexagonal lattice on the surface of a spherical vesicle. By virtue of their elastic properties~\cite{seung88}, these defects repel each other~\cite{bowick00} to arrange at the vertices of a platonic icosahedron. 

\begin{figure}[b]
\includegraphics{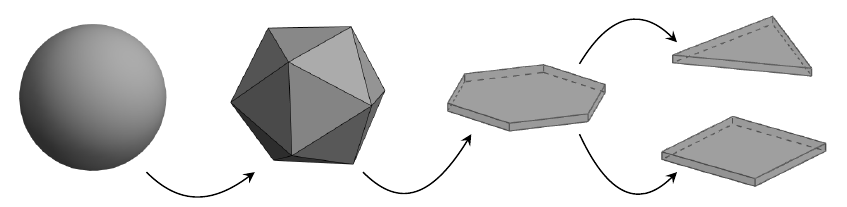}
\caption{Shape-Shifting Droplets. Main stages of the droplet shape evolution, following Refs.~\cite{denkov15,cholakova16}: the initially spherical droplets become icosahedral due to the interplay of topology and elasticity. Subsequently, the droplets flatten into hexagonal platelets which then evolve into triangles or quadrilaterals. The evolution of polygonal droplets was studied theoretically in Ref.~\cite{haas17}; the analysis of the flattening process is the subject of this paper.}
\label{fig:intro} 
\end{figure}

These same topological considerations play their part in the phenomenon of `shape-shifting' droplets reported by Denkov \emph{et al.}~\cite{denkov15}: micron-sized oil droplets in aqueous surfactant solution flatten, upon slow cooling, into a plethora of polygonal shapes with straight edges and sharp vertices \wholefigref{fig:intro}. Although first revealed briefly over a decade ago~\cite{*[] [{ and in particular p.~363 of the corresponding general discussion, }] sloutskin05,*fdiscuss,golemanov06}, these phenomena generated a veritable flurry of largely experimental papers~\cite{denkov15,denkov16,cholakova16,tcholakova17,cholakova17,valkova17,lesov18,guttman16a,guttman16b,guttman17,marin19,martin17} only more recently. These studies revealed that the shape-shifting phenomena arise for a humongous range of surfactants and pure organic phases or mixtures thereof~\cite{cholakova16,cholakova17}, and showed how to harness these phenomena for efficient, controlled self-emulsification~\cite{tcholakova17,valkova17}. More recent studies demonstrated their manufacturing potential by synthesizing small polymeric particles~\cite{lesov18,marin19}; scaled-up versions of these bottom-up approaches may enable massively parallel control over internally determined particle shape and particle uniformity that are currently only available in top-down approaches such as lithography techniques~\cite{lesov18,marin19}.

In spite of this large number of experimental studies and these manufacturing applications, the mechanisms underlying these phenomena remain debated, although there is agreement that the initial deformations of the droplets are caused by freezing of the surfactant adsorption layer~\cite{denkov15,denkov16,guttman16a,guttman16b,guttman17} and the ensuing topological frustration of the hexagonal packing of the surfactant molecules therein, leading to a transient icosahedral shape~\cite{guttman16a,guttman16b,guttman17}. Two driving mechanisms for the subsequent deformations have however been proposed: (i) a partial phase transition of the bulk oil phase~\cite{denkov15,denkov16,cholakova16,valkova17}, and (ii) elastic buckling of the frozen surfactant layer~\cite{guttman16a,guttman16b,guttman17}. According to the first mechanism, the formation of a plastic rotator phase~\cite{sirota99} of self-assembled oil molecules with long-range translational order becomes energetically favourable next to the droplet surface due to the freezing of the surfactant layer. This rotator phase then arranges into a scaffold of plastic rods at the surface of the droplet supporting the faceted droplet structure. We have shown in earlier theoretical work~\cite{haas17} that the rotator phase mechanism can account for the sequence of polygonal shapes seen in experiments, the statistics of shape outcomes and the observation that some droplets puncture in their centre before freezing~\cite{denkov15}. According to the second mechanism, elastic deformations can dominate over surface energy due to ultra-low values of surface tension resulting from the cooling.

\begin{figure*}
\includegraphics{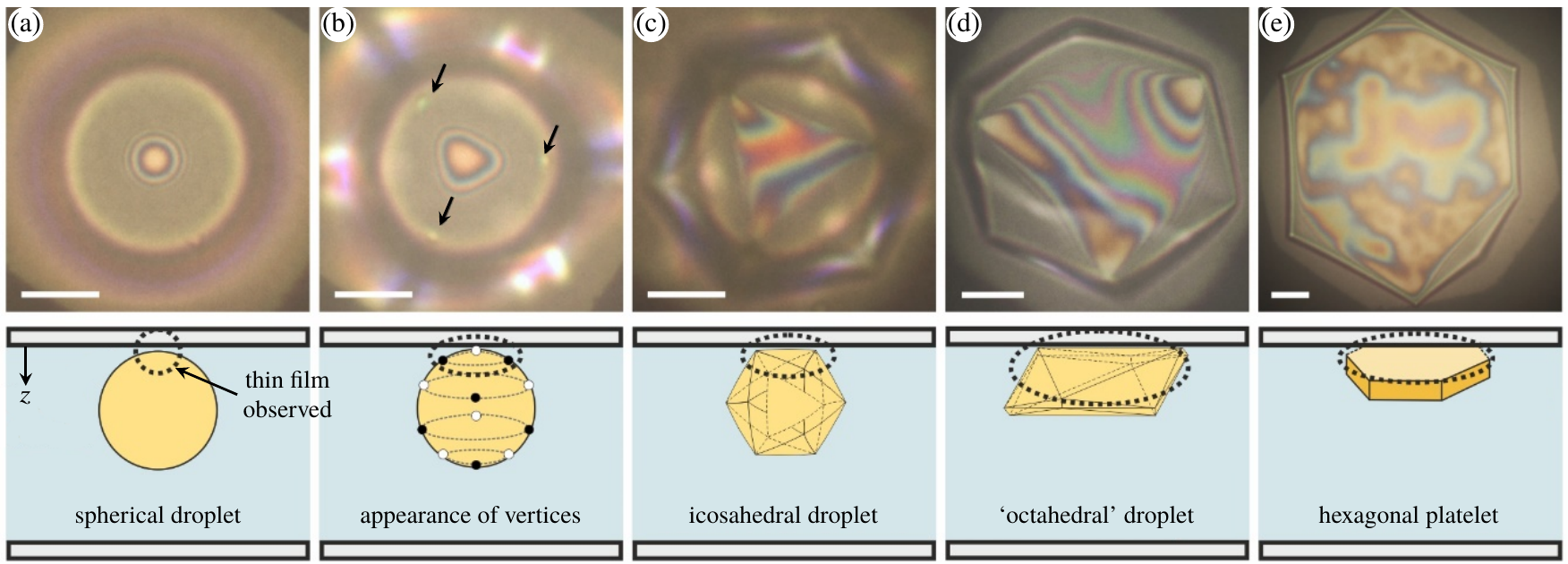}
\caption{Droplet shape evolution observed in reflected light. Top row: microscopy images obtained upon cooling of a hexadecane emulsion droplet immersed in $1.5\,\mathrm{wt.\%}$ Tween 60 surfactant solution; scale bars: $10\,\text{\textmu m}$. Bottom row: sketch of side view of droplet deformations. Dotted circles and ellipses show the part of the drop observed in reflected-light experiments. (a)~Before the drops start to deform, the aqueous film formed between the top of the drop and the wall of the glass capillary appears as circular Newton rings in reflected light. (b)~The emulsion film deforms into a triangular shape when the drop begins to deform. Black arrows show three of the 12 vertices formed on the drop surface at this stage. (c)~Upon further cooling, the triangular film expands until its corners engulf the three vertices at the periphery of the droplet that are situated above the equatorial plane of the icosahedral droplet. The three vertices forming the bottom surface of the drop appear as bright spots. (d)~As the flattening process continues, pairs of vertices of the icosahedron come closer to each other: the droplet becomes octahedral. (e)~As the droplet flattening completes, six pairs of vertices merge and the drop becomes a hexagonal platelet.}
\label{fig:exp1} 
\end{figure*}

The early three-dimensional polyhedral stages of the droplet evolution, from an initial icosahedron down to a flattened hexagon \wholefigref{fig:intro}, have so far remained unexplored, however. They are the subject of this paper: here, we analyse the flattening of an icosahedral droplet into a hexagonal platelet in detail, comparing experimental observations to predictions of mathematical models describing either mechanism to decide which mechanism underlies the observed phenomena. 

Thus, using reflected-light microscopy, we reveal how an icosahedral droplet flattens via an intermediate octahedral stage. Through a linear stability analysis and numerical calculations, we show that the rotator phase mechanism can reproduce the observed flattening dynamics for suitable choices of the microscopic law describing the formation of the rotator phase. The elastic buckling mechanism however can only reproduce the observed deformations if surface tension decreases by a least four orders of magnitude, so that the droplet evolution is driven by the interplay of elasticity and buoyancy. The analysis therefore suggests that it is formation of rotator phase rather than elastic buckling at low surface tension that drives the droplet shape evolution.

\section{Experimental Flattening Dynamics}
The flattening of the shape-shifting droplets under cooling and the stages of the droplet evolution intermediate between the initial spherical stage and the later flattened stages were observed using reflected-light microscopy to determine the three-dimensional shapes of the droplets at different stages of evolution. The experimental setup is described in Appendix~\ref{appA}.

\begin{figure}
\includegraphics{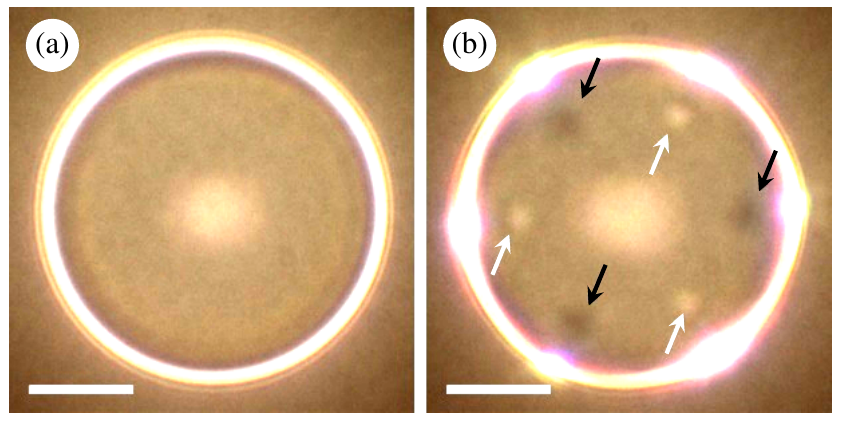}
\caption{Initial stages of the droplet shape deformations observed in reflected light with focus on the equatorial plane of the droplet. (a)~Before the droplet starts to deform, the droplet equator appears as a homogeneous bright circle. (b)~As the drop begins to deform, twelve vertices are observed: dark dots and bright spots (marked with black and white arrows) represent, respectively, the vertices at the top and bottom surfaces of the droplet. The remaining six vertices are located the drop periphery, just above and below the equatorial plane of the droplet. Scale bars: $10\,\text{\textmu m}$.}
\label{fig:exp2} 
\end{figure}

The droplets are initially spherical. Two well-defined types of images can be observed, depending on the position of the focal plane of the microscope: first, if the microscopy focus is on the top of the drop, just below the level of the upper wall of the glass capillary containing the emulsion, circular diffraction fringes (Newton rings) are seen \figref{fig:exp1}{a}. These fringes emerge from the interference of the light reflected from the two surfaces of the aqueous film, formed between the wall of the glass capillary and the surface of the spherical oil droplet \figref{fig:exp1}{a}. Second, if the microscope is focused on the equatorial plane of the droplet instead, a bright circle around the particle periphery is observed, due to light refraction and reflection at the drop surface \figref{fig:exp2}{a}. 

Deformation of the droplets begins with the appearance of twelve vertices on the drop surface [Figs.~\hyperref[fig:exp1]{\ref*{fig:exp1}(b)} and \hyperref[fig:exp2]{\ref*{fig:exp2}(b)}]. With the focal plane at the droplet equator, all twelve vertices can be observed simultaneously as three black dots (representing the three upper vertices next to the glass capillary), three white dots (representing the three vertices at the bottom of the droplet) and six bright spots at the drop periphery \figref{fig:exp2}{b}. The latter correspond to three vertices just above and three vertices just below the equatorial plane of the droplet, which explains why they have a slightly different appearance in the microscopy images of \textfigref{fig:exp2}{b}. As a result, the spherical shape of the drop is distorted and the droplet soon acquires an icosahedral shape; we note that this ideal shape transformation is, however, only observed in some of the systems under appropriate conditions such as slow cooling~\cite{denkov15,cholakova16}. At the same time, the aqueous film between the glass capillary and the oil droplet acquires an approximately triangular shape with rounded corners \figref{fig:exp1}{b}. Upon further cooling, the drop continues to deform so that this aqueous film appears as an equilateral triangular shape with sharp corners \figref{fig:exp1}{c}.

As cooling continues, this triangular film increases its area significantly. At the same time, the cross-section of the droplet equator also increases in size. Since the volume of the droplet is conserved, the droplet flattens in the perpendicular direction. Although the droplet looks like a hexagonal prism in transmitted light at this stage, the images in reflected light reveal that the three-dimensional drop shape is better represented as a distorted flattened icosahedron \figref{fig:exp1}{d}. The flattening of the icosahedral droplet drives pairs of vertices closer to each other, so that the droplet assumes the shape of a flattened octahedron. Eventually, pairs of vertices can merge to form true hexagonal platelets \figref{fig:exp1}{e}, but the details of this final step depend on the system (Appendix~\ref{appA}).

To understand these complex droplet shape deformations, we derive theoretical models corresponding to the two proposed mechanisms~\cite{denkov15,denkov16,cholakova16,guttman16a,guttman16b,guttman17} in the next section.

\section{Model}
On purely combinatorial grounds, the appearance of octahedral droplets as the vertices of the initial icosahedron merge during the flattening process is not suprising: indeed, the octahedron is one of only two polyhedra with six vertices that can be obtained by edge contraction from an icosahedron, and the only one that does not require additional symmetry breaking (Appendix~\ref{appB}). Static, entropic considerations of this ilk cannot however capture the dynamics of the problem: as in our previous theoretical description of the dynamics of flattened polygonal droplets~\cite{haas17}, describing the deformations of a polyhedron requires (i) a (non-dimensional) energy $\mathcal{E}$, and (ii) a kinetic law that relates the variations of the energy to the normal velocity of the edges of the polyhedron. 

In this paper, we model the polyhedral droplets as convex polyhedra of fixed volume, with flat faces. Throughout the paper, we will use $\mathscr{E}$ and $\mathscr{F}$ to denote, respectively, the set of edges and faces of such a polyhedral droplet.

\subsection{Droplet Energy}
In this section, we derive different energies describing the two mechanisms that have been proposed to underlie the shape-shifting phenomena~\cite{denkov15,denkov16,cholakova16,guttman16a,guttman16b,guttman17}.

\subsubsection{Rotator phase mechanism}
For the rotator-phase mechanism, the energy has contributions from surface tension and from the rotator phase. Extending the energy we have derived to describe the polygonal stages of the droplets~\cite{haas17} to the polyhedral case,
\begin{align}
\mathcal{E}_1=\sum_{f\in\mathscr{F}}{\|f\|}-\alpha\sum_{e\in\mathscr{E}}{\|e\|F\bigl(\delta(e)\bigr)},  \label{eq:E1}
\end{align}
wherein, as in our previous work~\cite{haas17}, the coefficient $\alpha$ depends on temperature and has the scaling $\alpha\sim A_{\mathrm{r}}\upDelta\mu/\gamma R$, in which $A_{\mathrm{r}}$ is a characteristic cross-sectional area of the rotator phase, $\upDelta\mu=\mu_{\mathrm{l}}-\mu_{\mathrm{r}}>0$ is the difference of the chemical potentials (per unit volume) of the liquid and rotator phases, $\gamma$ is the coefficient of surface tension and $R\approx 10\,\text{\textmu m}$ is a typical radius of the shape-shifting droplets~\cite{denkov15}.

The function $F$ encodes the dependence of the formation of rotator phase on the dihedral angle $\delta(e)$ at edge $e$. This function did not arise in our previous work~\cite{haas17}, for it is constant during the polygonal stages of the shape evolution. The detailed functional form of $F$ is set by the microscopic properties of the rotator phase, but we expect $F$ to be a decreasing function of $\delta(e)$, with
\begin{align}
F(0)\geq F\bigl(\delta(e)\bigr)\geq F(\pi)=0, \label{eq:fconds}
\end{align}
so that no rotator phase is formed when the two faces adjacent to edge $e$ are parallel to each other, while the tendency to form rotator phase is maximal when the two faces have folded on top of each other.

\subsubsection{Elastic buckling of a frozen monolayer}
For the elastic buckling mechanism, the energy has a first contribution from surface tension, and a second contribution from surface elasticity. In the regime of sharp elastic ridges relevant to our discussion, the elastic energy is dominated by the contributions from the ridges~\cite{lidmar03}. Invoking the scaling properties of these elastic ridges derived in Ref.~\cite{lobkovksy96},
\begin{align}
\mathcal{E}_2=\sum_{f\in\mathscr{F}}{\|f\|}+\beta\sum_{e\in\mathscr{E}}{\|e\|^{1/3}\bigl(\pi-\delta(e)\bigr)^{7/3}}, \label{eq:E2}
\end{align}
where $\delta(e)$ again denotes the dihedral angle at edge $e$, and wherein the coefficient $\beta$ depends on temperature and has the scaling $\beta\sim Eh^{8/3}/\gamma R^{5/3}$, in which $E$ is the elastic modulus of the frozen surfactant adsorption layer, $h$ is its thickness, and, as before, $\gamma$ is the coefficient of surface tension and $R\approx 10\,\text{\textmu m}$ is a typical radius of the shape-shifting droplets. 

Upon including an additional buoyancy term $\mathcal{B}$, this energy becomes instead
\begin{align}
\mathcal{E}_2'=\mathcal{E}_2+\mathrm{Bo}\,\mathcal{B}, 
\end{align}
wherein the non-dimensional Bond number $\mathrm{Bo}=\upDelta\rho\,g R^2\big/\gamma$ measures the relative magnitude of buoyancy and surface tension effects~\cite{bond}. Here, $\upDelta\rho\approx 250\,\mathrm{kg/m^3}$ is the density difference between the water and oil phases~\cite{*[{At standard conditions, alkanes with 14--20 carbon atoms have densities of about $750$ kg/m$^3$, see e.g. }] [] crc}, and $g\approx 9.81\,\mathrm{m/s^2}$ is the acceleration due to gravity.

There is no consensus on the value of $\gamma$ at the temperature $T_{\mathrm{d}}$ at which the deformations are first observed~\cite{denkov16,cholakova16,guttman17}, and the proponents of the elastic buckling mechanism indeed argue that the observed deformations are driven by ultra-low values of $\gamma$~\cite{guttman16a,guttman16b,guttman17}. Nevertheless, using the estimate $\gamma\approx5\,\mathrm{mN/m}$ above $T_{\mathrm{d}}$ that the different experimental analyses ~\cite{denkov16,cholakova16,guttman17} agree on, we estimate $\mathrm{Bo}\approx 10^{-4}$, and conclude that $\gamma$ must decrease by at least four orders of magnitude during the cooling for buoyancy effects to play a role.

Using the rotator phase mechanism, we have previously described the platelet stages of the droplet evolution by modelling the droplets as polygons~\cite{haas17}. There, disproportionation of the sidelengths of a polygonal droplet of fixed area is energetically favourable because, among all polygons of a fixed number of vertices and fixed area, the regular one has the least perimeter~\cite{*[{Consider indeed a regular $n$-gon of area $A_0$ and perimeter $P_0$, and suppose to the contrary that there is a $n$-gon of area $A_0$ and perimeter $P_1<P_0$. Shrinking the regular $n$-gon, there thus exist a regular $n$-gon of area $A_1<A_0$ and perimeter $P_1$, and a $n$-gon of area $A_0$ and perimeter $P_1$. This is a contradiction, for among all $n$-gons of fixed perimeter, the regular $n$-gon has the largest area, see e.g. }] [] tikhomirov}. By contrast, in the elastic buckling mechanism, disproportionation of the sidelengths of a polygonal droplet is energetically favourable because the second term in Eq.~(\ref{eq:E2}) is a concave function of edge length~\cite{*[{Jensen's inequality [see e.g. }] [{] implies that the elastic energy of a non-regular polygon is lower than that of a regular polygon of the same perimeter, as the elastic energy is a concave function of edge length. However, this does not imply that disproportionation of sidelengths is energetically favourable since a non-regular polygon of the same area must have a larger perimeter than the original regular polygon. That disproportionation is indeed favourable can be proved by a linear stability analysis similar to \S\ref{linstab}} which therefore confirms the qualitative convexity argument.] stirzaker}. 

\subsection{Kinetic Law}
\begin{figure}[b]
\includegraphics{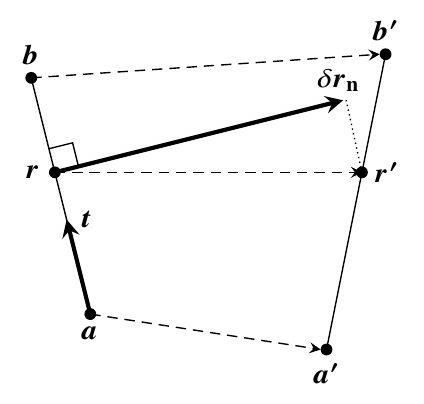} 
\caption{Derivation of the kinetic law: definition of the normal displacement $\delta\vec{r_{\mathrm{n}}}$ of a point $\vec{r}$ on the edge parallel to $\vec{t}$ and joining the vertices at $\vec{a}$ and $\vec{b}$ that move to $\vec{a'}$ and $\vec{b'}$ as the polyhedron deforms.}
\label{fig:kingeo}
\end{figure}

In our previous theoretical description of the polygonal stages of the droplet evolution~\cite{haas17}, we imposed a simple kinetic law~\cite{suo97}, that the normal velocity of the sides be proportional to the energy gradient. The three- and non-dimensional analogue of this kinetic law is
\begin{align}
-\delta\mathcal{E}=\sum_{e\in\mathscr{E}}{\int_e{\vec{\dot{r}_{\mathrm{n}}}\cdot\delta\vec{r_{\mathrm{n}}}\,\mathrm{d}\ell}}. 
\end{align}
wherein $\vec{r_{\mathrm{n}}}$ is the normal displacement of edge $e$ of the polyhedron. Consider the edge joining vertices $\vec{a}$ and $\vec{b}$, and let $\vec{t}$ denote the unit tangent parallel to it \wholefigref{fig:kingeo}. As the polyhedron deforms from its initial shape, described by some variables $\vec{x}$, to a new shape described by $\vec{x}+\delta\vec{x}$, these vertices are mapped to $\vec{a'}=\vec{a}+\mat{A}\cdot\delta\vec{x}$ and $\vec{b'}=\vec{b}+\mat{B}\cdot\delta\vec{x}$. Hence, assuming that the edges stretch uniformly, a point $\vec{r}=\vec{a}+(\vec{b}-\vec{a})s$, where $0\leq s\leq 1$, is mapped to $\vec{r}+\mat{R}\cdot\delta\vec{x}$, with $\mat{R}=\mat{A}+(\mat{B}-\mat{A})s$, as shown in \textwholefigref{fig:kingeo}. 

Let $\mat{P}=\mat{I}-\vec{t}\vec{t}$ denote projection onto the plane normal to~$\vec{t}$. Then $\delta\vec{r_{\mathrm{n}}}=\mat{P}\cdot\mat{R}\cdot\delta\vec{x}$ and so $\vec{\dot{r}_{\mathrm{n}}}=\mat{P}\cdot\mat{R}\cdot\vec{\dot{x}}$. Using $\mat{P}^2=\mat{P}$, it follows that
\begin{align}
\int_e{\vec{\dot{r}_{\mathrm{n}}}\cdot\delta\vec{r_{\mathrm{n}}}\,\mathrm{d}\ell} =\|e\|P_{ij}\delta x_k\dot{x}_\ell\int_0^1{R_{ik}R_{j\ell}\,\mathrm{d}s}, 
\end{align}
with, upon letting $\mat{S}=\mat{A}+\mat{B}$,
\begin{align}
\int_0^1{R_{ik}R_{j\ell}\,\mathrm{d}s}=\dfrac{1}{6}\Bigl(S_{ik}S_{j\ell}+A_{ik}A_{j\ell}+B_{ik}B_{j\ell}\Bigr).
\end{align}
The factor $1/6$ in this equation can and will be scaled out by rescaling time. Thence
\begin{align}
-\dfrac{\partial\mathcal{E}}{\partial x_k}\delta x_k=\dot{x}_\ell\delta x_k\sum_{e\in\mathscr{E}}{\|e\|P_{ij}\Bigl(S_{ik}S_{j\ell}+A_{ik}A_{j\ell}+B_{ik}B_{j\ell}\Bigr)}, 
\end{align}
which leads to the overdamped evolution equation
\begin{align}
\vec{\dot{x}}=-\mat{M}^{-1}\vec{\nabla}\mathcal{E},\label{eq:goveq}
\end{align}
wherein 
\begin{align}
\mat{M}=\sum_{e\in\mathscr{E}}{\|e\|\Bigl(\mat{S}^\top\mat{PS}+\mat{A}^\top\mat{PA}+\mat{B}^\top\mat{PB}\Bigr)}.\label{eq:M}
\end{align}
We shall refer to $\mat{M}$ as the mobility matrix. A more standard kinetic law~\cite{suo97} would instead involve the normal velocity of the faces being proportional to the energy gradient. We have not chosen this kinetic law, since the motion is driven by the edges of the polyhedron, rather than by its faces. Nevertheless, the equations corresponding to this second kinetic law can be derived by analogous reasoning, and we have checked that choosing this kinetic law yields qualitatively similar results.

\section{Results}

\subsection{Linear Stability Analysis}\label{linstab}
A necessary (albeit not sufficient) condition for droplet flattening is that the initial regular icosahedron be unstable to small perturbations. For the linear stability analysis, we describe the icosahedron by means of the coordinates $\vec{x}=(\vec{x_1},\dots,\vec{x_{12}})$ of its vertices and introduce the Lagrangian
\begin{align}
\mathcal{L}=\mathcal{E}-\lambda\mathcal{V}, \label{eq:L}
\end{align}
wherein 
\begin{align}
\mathcal{V}=\hspace{-3mm}\sum_{(k,\ell,m)\in\mathscr{F}}{\hspace{-3mm}\bigl|\vec{x_k}\cdot(\vec{x_\ell}\times\vec{x_m})\bigr|} 
\end{align}
is the volume of the icosahedron, and $\lambda$ is the Lagrange multiplier imposing volume conservation. Let $\vec{x}^\ast$ denote the coordinates of the platonic icosahedron; imposing $\vec{\nabla}\mathcal{L}\bigl(\vec{x}^\ast,\lambda^\ast\bigr)=\vec{0}$ yields the corresponding value $\lambda^\ast$ of the Lagrange multiplier. The Hessian for this stability problem is $\mat{H}=\mat{P}\bigl(\vec{\nabla\nabla}\mathcal{L}\bigr)\mat{P}$, where the matrix $\mat{P}=\mat{I}-\vec{vv}$ describes the projection onto the kernel of $\vec{v}=\vec{\nabla}\mathcal{V}\bigl(\vec{x}^\ast,\lambda^\ast\bigr)$~\cite{optim}. We note in passing that, since the mobility matrix is invertible and therefore an isomorphism, it does not affect the stability analysis.

\subsubsection{Rotator phase mechanism}
For each face $(k,\ell,m)\in\mathscr{F}$, we define
\begin{align}
\vec{n_{k\ell m}}=\pm\Bigl(\vec{x_k}\times\vec{x_\ell}+\vec{x_\ell}\times\vec{x_m}+\vec{x_m}\times\vec{x_k}\Bigr)
\end{align}
to be its outward normal, so that Eq.~(\ref{eq:E1}) becomes
\begin{align}
\mathcal{E}_1&=\hspace{-3mm}\sum_{(k,\ell,m)\in\mathscr{F}}{\hspace{-3mm}\tfrac{1}{2}\|\vec{n_{k\ell m}}\|}-\alpha\hspace{-3mm}\sum_{\substack{(k,\ell,m)\in\mathscr{F}\\(k,\ell,n)\in\mathscr{F}}}{\hspace{-3mm}\|\vec{x_k}-\vec{x_\ell}\|F(\delta_{k\ell mn})},\label{eq:E1b}
\end{align}
wherein $\delta_{k\ell mn}$ is defined by
\begin{align}
\cos{\delta_{k\ell m n}}=\dfrac{\vec{n_{k\ell m}}\cdot\vec{n_{k\ell n}}}{\left\|\vec{n_{k\ell m}}\right\|\left\|\vec{n_{k\ell n}}\right\|}.
\end{align}
Formally, we may write
\begin{align}
\mat{H}^\ast=\mat{S}^\ast-\alpha\Bigl(F(\delta^\ast)\mat{A^\ast}+F'(\delta^\ast)\mat{B^\ast}+F''(\delta^\ast)\mat{C^\ast}\Bigr), \label{eq:Hess}
\end{align}
where the matrices $\mat{A^\ast},\mat{B^\ast},\mat{C^\ast},\mat{S}^\ast$ are purely geometric, and where $\delta^\ast\approx 138.2^\circ$ is the dihedral angle of the platonic icosahedron. We set $F(\delta^\ast)=1$ without loss of generality. We sketch the calculations leading to expressions for these matrices in Appendix~\ref{appC}. 

\begin{figure}[t]
\includegraphics{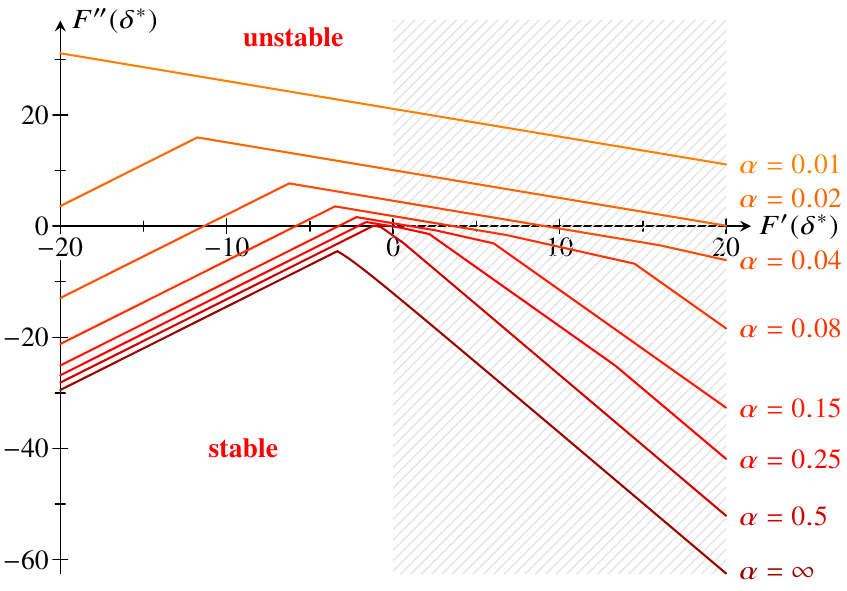}
\caption{Stability of the platonic icosahedron for the rotator phase mechanism. The stability boundary is shown in $\bigl(F'(\delta^\ast),F''(\delta^\ast)\bigr)$ space for different values of $\alpha>0$. Only the region with $F'(\delta^\ast)<0$ is expected to be physically relevant.}
\label{fig:stab} 
\end{figure}

Evaluating these expressions numerically~\cite{SM} using \textsc{Matlab} (The MathWorks, Inc.), we find in particular that $\mat{S^\ast}$ is positive semi-definite, so the platonic icosahedron is stable if $\alpha=0$, as expected. The matrices $\mat{A^\ast},\mat{B^\ast}$ are indefinite, but $\mat{C^\ast}$ is positive semi-definite. It follows that, for $\alpha>0$, a regular icosahedron is unstable provided that $F''(\delta^\ast)$ is large enough. More generally, the stability boundary for $\alpha>0$ can be computed numerically by a bisection search; results are shown in \textwholefigref{fig:stab}. Kinks in the curves defining the stability boundary indicate different eigenvalues crossing zero at the stability boundary. We conclude that a regular icosahedron is unstable to small perturbations for appropriate choices of the microscopic law~$F$. For a fixed choice of microscopic law, only droplets that are small enough (i.e. have large enough values of $\alpha$) deform; larger droplets are stable \wholefigref{fig:stab}. The rotator phase mechanism can thus explain the deformations away from the initial platonic icosahedron. 

The above analysis can be extended to any platonic solid. In the particular case of a regular octahedron, the geometry of the eigenmodes is much simpler, and we therefore discuss these eigenmodes briefly in Appendix~\ref{appD}.

\subsubsection{Elastic buckling mechanism}
For the elastic buckling mechanism (in the absence of the buoyancy term), starting from Eq.~(\ref{eq:E2}), a similar calculation leads to the expression $\mat{H}^\ast= \mat{S}^\ast+\beta\mat{D}^\ast$, in which the purely geometric matrices $\mat{D}^\ast,\mat{S}^\ast$ are found to be positive semi-definite upon numerical evaluation~\cite{SM}, and hence $\mat{H}^\ast$ is positive semi-definite, too, for any $\beta\geq 0$, since the sum of positive semi-definite matrices is positive semi-definite. The key conclusion from the stability analysis is therefore that the elastic buckling mechanism cannot explain the deformations of the initial icosahedron unless buoyancy effects become important.

\subsection{Droplet Flattening}
The linear stability analysis has revealed necessary conditions for the initial regular icosahedron to deform under either mechanism. Larger deformations, and, in particular, flattening, of the polyhedral droplets must however be studied numerically.

\begin{figure}[b]
\includegraphics{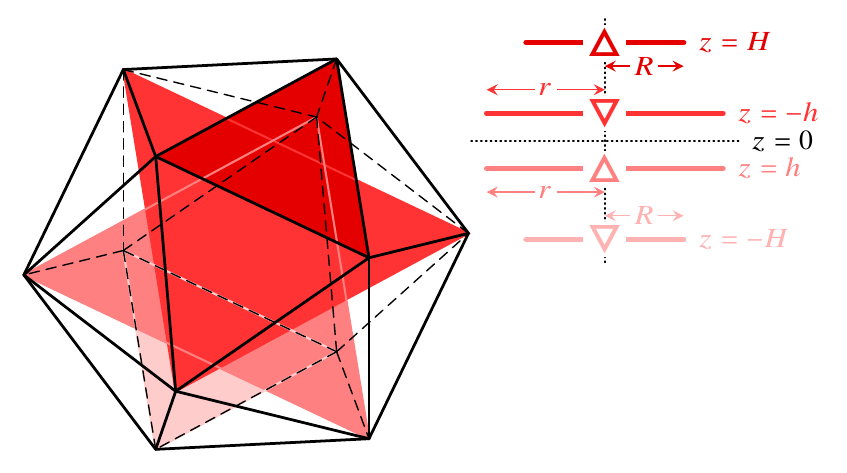}
\caption{Simplified model of a symmetric icosahedron. Four parallel equilateral triangles that are symmetric with respect to the midplane of the polyhedron define a symmetric icosahedron in terms of their circumradii $r,R$ and vertical positions $\pm h,\pm H$. Inset shows definitions of variables $r,R,h,H$, and orientations of the four equilateral triangles.}
\label{fig:icosdef}
\end{figure}

\begin{figure*}
\includegraphics{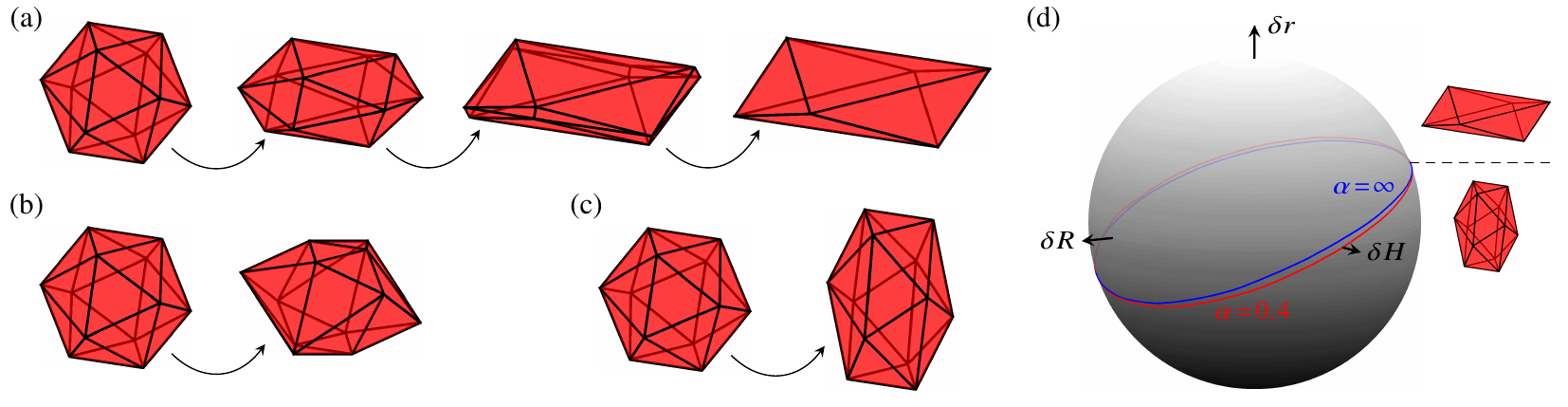}
\caption{Flattening of a symmetric icosahedron under the rotator phase mechanism. (a)~Snapshots of a slightly perturbed platonic icosahedron flattening into a (non-platonic) octahedron, for $\alpha=\infty$, and for the microscopic law $F(\delta)\propto\pi^3-\delta^3$. (b)~Role of the microscopic law: for a different microscopic law, here $F(\delta)\propto\pi-\delta$, but the same initial perturbation, the icosahedron does not flatten into an octahedron. (c)~Role of the initial perturbation: for the same microscopic law, but a different initial perturbation, the icosahedron does not flatten either. (d)~Three of the four parameters defining the symmetric icosahedron \wholefigrefi{fig:icosdef} can be chosen as a basis for perturbations away from the regular icosahedron. Shape outcomes for such perturbations of fixed magnitude $\varepsilon=0.02$ are then mapped onto the surface of a sphere. The boundary between those initial relative perturbations leading to flattening into an octahedron and those that do not lead to flattening are shown for $F(\delta)\propto\pi^3-\delta^3$ and different values of $\alpha$.}
\label{fig:flattrot}
\end{figure*}

The experimental data suggest that the icosahedron flattens symmetrically \wholefigref{fig:exp1}, and hence that the four parallel equilateral triangles that define a platonic icosahedron \wholefigref{fig:icosdef} remain equilateral during the flattening, although their relative positions with respect to the flattening axis changes. (It is only after the icosahedron has flattened that this symmetry broken as the polygonal droplets deform. In other words, the component of the initial perturbations of the icosahedron corresponding to these asymmetric deformations is so small that it remains small during the entire flattening process.) This suggests using a simplified representation of a symmetric icosahedron, defined by four parallel equilateral triangles \wholefigref{fig:icosdef} for the numerical calculations. Such an icosahedron is defined in terms of four variables \wholefigrefi{fig:icosdef}; one of these can be eliminated using the volume conservation constraint (Appendix~\ref{appE}). 

We solve equation (\ref{eq:goveq}) governing the deformations of the icosahedron numerically using the stiff solver \texttt{ode15s} of \textsc{Matlab} (The MathWorks, Inc.).

\subsubsection{Rotator phase mechanism}
We begin by considering the rotator phase mechanism and the limit $\alpha=\infty$ where the tendency to form rotator phase swamps the stabilising effect of surface tension. Our first observation is that a (slightly perturbed platonic) symmetric icosahedron may indeed flatten into an octahedron \figref{fig:flattrot}{a} under the rotator phase mechanism: as the icosahedron flattens, the top and bottom equilateral triangles expand faster than the middle ones, leading to six pairs of vertices merging to yield a (non-platonic) octahedron, in qualitative agreement with the shape evolution seen in experiments. This evolution depends on the choice of the microscopic law $F(\delta)$ and the initial perturbation. As far as the choice of $F(\delta)$ is concerned, flattening occurs for example for $F(\delta)\propto \pi^3-\delta^3$ \figref{fig:flattrot}{a}, but does not occur for the simplest law in agreement with conditions~(\ref{eq:fconds}), $F(\delta)\propto\pi-\delta$ \figref{fig:flattrot}{b}. We have checked that the behaviour in \textfigref{fig:flattrot}{a} is representative of the behaviour observed for sufficiently concave choices of the microscopic law $F(\delta)$ that verify conditions~(\ref{eq:fconds}). As far as the initial perturbations are concerned, flattening similarly occurs for some, but not all perturbations of the regular icosahedron [Fig.~\hyperref[fig:flattrot]{\ref*{fig:flattrot}(a)},\hyperref[fig:flattrot]{(c)}]. 

To explore the latter effect and the role of surface tension, we consider relative perturbations, of fixed magnitude $\varepsilon$, of the parameters defining the symmetric icosahedron \wholefigrefi{fig:icosdef}. Taking three of these parameters as the basis for the perturbations without loss of generality, we map the shape outcomes for the corresponding initial perturbations onto the surface of a sphere \figref{fig:flattrot}{d}. A boundary divides those initial perturbations that lead to flattening to those that do not, and we observe that this boundary does not strongly depend on~$\alpha$ \figref{fig:flattrot}{d}. Physically, the initial perturbation is set by the buoyancy of the droplets. Using the coordinates $r,R,h$ defined in the inset of \textwholefigref{fig:icosdef}, we expect the physically relevant initial perturbations of the droplets to be those with $\delta H<0$ and $\delta r,\delta R>0$, for which the droplet indeed flattens into an octahedron.

We conclude that, for an appropriate choice of microscopic law, our model of the rotator phase mechanism predicts flattening in qualitative agreement with the experimental observations for physical initial perturbations of the icosahedron. 

\begin{figure}[b]
\includegraphics{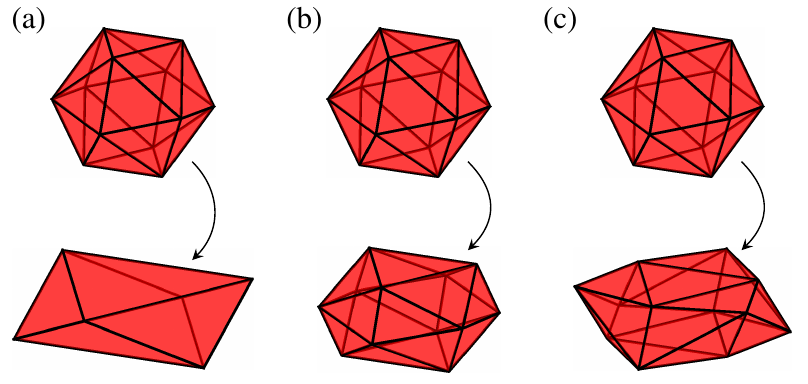}
\caption{Flattening of a symmetric icosahedron under the elastic buckling mechanism for different values of $B=\mathrm{Bo}/\beta$. (a)~Flattening of a platonic icosahedron into a (non-platonic) octahedron. (b)~If $B$ is too small, the icosahedron settles into an unflattened steady state. (c)~If $B$ is too large, the faces connected to the top face flatten into its plane, inconsistent with experimental observations.}
\label{fig:flattbuck}
\end{figure}

\subsubsection{Elastic buckling mechanism}
The stability analysis in the previous section has revealed that the regular icosahedron is a stable fixed point if buoyancy does not play a role, $\mathrm{Bo}=0$. We therefore consider the case $\mathrm{Bo}>0$, in which the regular icosahedron is therefore no longer a fixed point of the energy~$\mathcal{E}_2'$. We begin by analysing the limit of low surface tension, in which $\beta,\mathrm{Bo}\gg 1$ and the dynamics depend on the single parameter $B=\mathrm{Bo}/\beta$. Numerically, we find that there is an intermediate range, $B_-<B<B_+$, in which the platonic icosahedron flattens into an octahedron \figref{fig:flattbuck}{a}. We estimate $B_-\approx 7$ and $B_+\approx 140$. If $B<B_-$, the droplet settles into a steady state before reaching an octahedral shape \figref{fig:flattbuck}{b}. If $B>B_+$, the three faces adjacent to the top face flatten into its plane \figref{fig:flattbuck}{c}. Thus, the droplet evolves into a hexagonal prism as the top triangle continues to expand. This is inconsistent with the experimental observation that the top plane remains triangular \figref{fig:exp1}{c}, and does not become hexagonal until the end of the flattening.

Qualitatively similar results are obtained at non-zero surface tension; because of its stabilising effect, $B_-$ increases with increasing surface tension. We conclude that the elastic buckling mechanism predicts flattening in qualitative agreement with the experimental observations if buoyancy effects are neither too strong nor too weak.

To obtain more quantitative estimates, we notice the scaling $B\sim \upDelta\rho\,R^{11/3}\big/Eh^{8/3}$, wherein, as before, $E$ is the elastic modulus of the frozen surfactant monolayer and $h$ is its thickness, $g\approx 9.81\,\mathrm{m/s^2}$ is the acceleration due to gravity, $R$ is the droplet radius, and $\upDelta\rho\approx 250\,\mathrm{kg/m^3}$ is the density difference between the water and oil phases. Hence the bending modulus of the frozen surfactant layer is
\begin{align}
K=Eh^3\sim \dfrac{\upDelta\rho\,gR^{11/3}h^{1/3}}{B}. \label{eq:K}
\end{align}
Usefully, this allows us to obtain an upper bound on $K$ without having to estimate the surface tension, which does not appear in this expression: denote by $R_-$ the radius, corresponding to $B_-$, of the smallest droplet that can flatten; previous work~\cite{denkov15} has shown $R_-<5\,\text{\textmu m}$. Taking $h\approx 2\,\mathrm{nm}$~\cite{sloutskin07}, we thus obtain that $K<2\cdot 10^{-20}\,\mathrm{J}$ is required for flattening into an octahedron. By contrast, direct measurements of the bending moduli of shape-shifting droplets in Ref.~\cite{guttman16a} led to the lower bound $K>10^3\,\mathrm{k_BT}\gtrsim3\cdot 10^{-18}\,\mathrm{J}$, more than two orders of magnitude above the present upper bound. 

\section{Conclusion}
In this paper, we have analysed the flattening of shape-shifting droplets experimentally and theoretically. Models of the two candidate mechanisms have reproduced the evolution of an icosahedral droplet into a flattened octahedral shape in qualitative agreement with the experimental observations. The elastic buckling mechanism, however, can only reproduce the experimental observations if surface tension decreases by at least four orders of magnitude during the cooling, so that the flattening is driven by a competition between buoyancy and elasticity. Moreover, the resulting estimate of the bending modulus of the surfactant adsorption layer is two orders of magnitude too low. All of this strongly indicates that the `shape-shifting' droplet phenomena are driven by formation of rotator phase rather than elastic buckling of the frozen surfactant adsorption layer at ultra-low surface tension.

While the simple models used in this paper to represent the droplets as true polyhedra could thus reproduce the experimental flattening dynamics qualitatively, it is important to recognise that the faces of the actual droplets do not remain flat, but deform due to elasticity and surface tension. Similarly, these simple models do not take into account the dynamics of elastic defects in the surface. Repeating the computations in this paper for deformable faceted elastic surfaces with defects remains a formidable numerical challenge. Fully resolving the defect energetics would also make possible a more detailed analysis of the merging of the defects at the very latest stages of the flattening.

Some questions more specific to the rotator phase mechanism also remain open: in particular, how does the dependence of the energy on the dihedral angle, encoded in the functional form of $F(\delta)$, relate to properties of the rotator phase and the phase change? Deriving this microscopic law from first principles would remove some of the arbitrariness in the present analysis, where we chose a functional form \emph{ad hoc} for lack of knowledge about the detailed physics involved, even though, as noted previously, the qualitative behaviour of the model does not depend strongly on these details. The impact of the kinetics of phase change on the droplet evolution also remains unclear.

\begin{acknowledgments}
We thank Slavka Tcholakova for numerous fruitful discussions and for helping with the analysis of the microscopy data. We also thank Ireth Garc\'ia-Aguilar for a discussion of her unpublished work~\cite{aguilar} on shape-shifting droplets, and are grateful for support from the European Research Council (Grant EMATTER 280078 to S.K.S.), the Engineering and Physical Sciences Research Council (Established Career Fellowship EP/M017982/1 to R.E.G.; Fellowship EP/R028915/1 to S.K.S.; Doctoral Prize Fellowship to P.A.H.), the Schlumberger Chair Fund, Wellcome Trust Investigator Award 207510/Z/17/Z (R.E.G.), and Magdalene College, Cambridge (Nevile Research Fellowship to P.A.H.).
\end{acknowledgments}

\begin{figure}[t]
\includegraphics{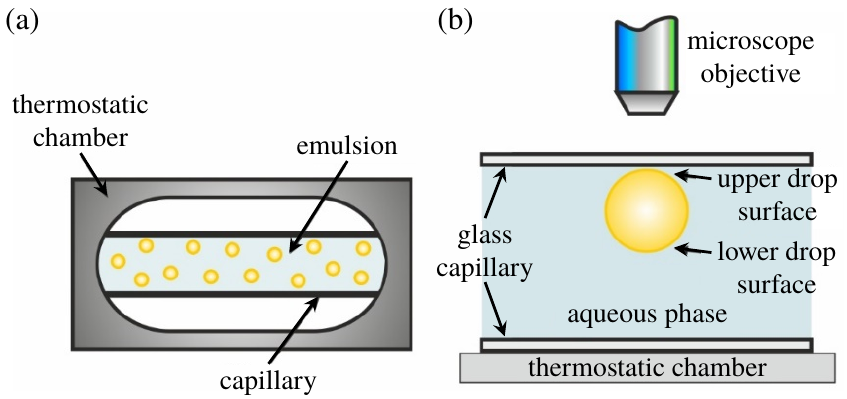}
\caption{Experimental setup for reflected-light microscopy of droplet deformations. (a)~Emulsions are examined in a glass capillary placed inside a thermostatic chamber for microscopic observations during cooling. (b)~Side view: due to buoyancy, oil droplets float just below the top surface of the capillary, under the objective of the microscope.}
\label{fig:expsetup}
\end{figure}

\begin{figure*}
\includegraphics{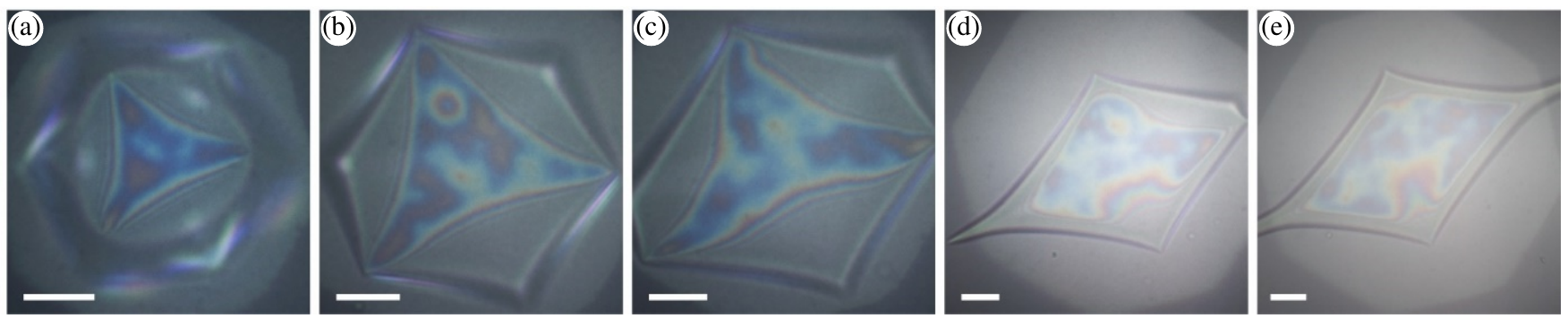}
\caption{Merging of the vertices of a polyhedral drop with formation of asperities. This experiment was performed with a hexadecane emulsion droplet, immersed in $1.5\,\mathrm{wt.\%}$ Brij 58 surfactant solution. Merging of vertices causes the octahedral stage (panel b) to disproportionate (panel~c), leading to the formation of a tetragonal platelet with asperities (panel e). Scale bars: $10\,\text{\textmu m}$.}
\label{fig:expvm}
\end{figure*}

\begin{figure*}
\includegraphics{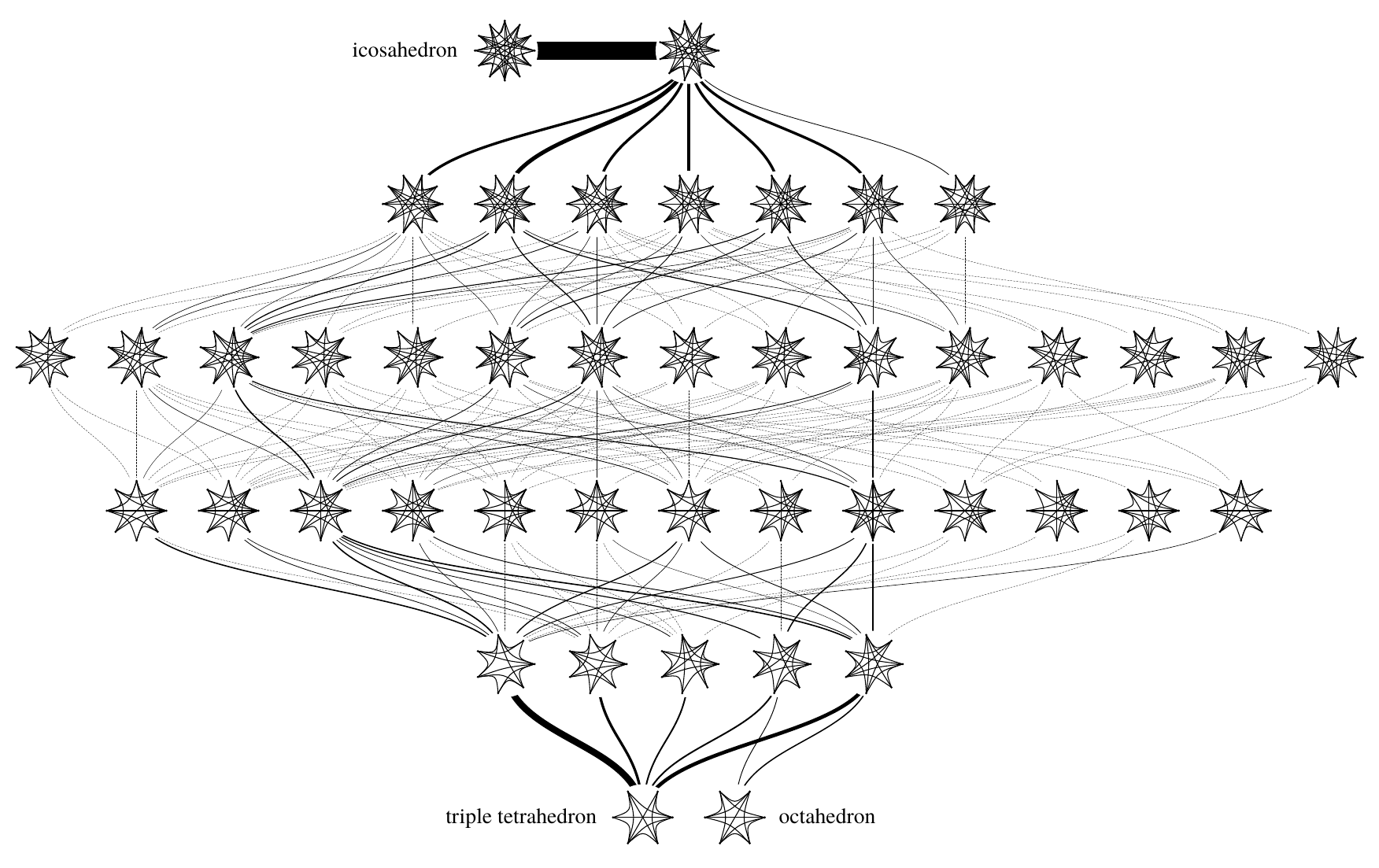}
\caption{Classification of polyhedra of a least six vertices obtained by edge contraction from an icosahedron. Edge graphs of the 44 polyhedra obtained are shown, with lines between polyhedra indicating possible edge contraction paths. Line widths are proportional to the probability that a random walk starting from the icosahedron passes through a particular transition. Small probabilities are represented by dotted lines. The icosahedron, octahedron, and triple tetrahedron are labelled.}
\label{fig:polyhedra}
\end{figure*}

\appendix\section{Experimental Methods and Details}\label{appA}
Oil emulsion droplets were observed in glass capillaries placed inside a thermostatic chamber below the objective of the microscope \wholefigref{fig:expsetup}, as described previously~\cite{denkov15,cholakova16}. Observations were performed with an upright optical microscope in reflected white light. Due to the buoyancy force, oil droplets float just below the upper wall of the glass capillary \figref{fig:expsetup}{b}. The aqueous film formed between this wall and the upper surface (closest to the capillary) of the droplet was observed in reflected light.

As mentioned in the main text, the details of the final step of the droplet flattening depend on the emulsion system: for those systems in which no asperities (long cylindrical protrusions from the platelet corners) are formed, i.e. in Group B systems according to the classification of Ref.~\cite{cholakova16}, pairs of vertices eventually merge to form true hexagonal platelets \figref{fig:exp1}{e}. By contrast, in those systems in which asperities do form, i.e. in Groups A and C of Ref.~\cite{cholakova16}, the merging of the vertices also drives the transition from an octahedral droplet into tetragonal or triangular plates \wholefigref{fig:expvm}. 

\section{Classification of a Family of Polyhedra}\label{appB}
In this appendix, we classify polyhedra that can be obtained by edge contraction from an icosahedron. Starting from an icosahedron, we thus contract edges to reduce the number of vertices of the polyhedron. Representing each polyhedron by its undirected edge graph $\mathscr{G}=(\mathscr{V},\mathscr{E})$ of vertices $\mathscr{V}$ and edges $\mathscr{E}$, we classify, numerically and up to isomorphism, the 105 graphs of at least six vertices obtained by edge contraction in this way. (Such a classification is of course a hard problem in general, but the graphs are small enough for classification by brute force to be straightforward.) Requiring these graphs to correspond to true polyhedra, we require that any edge be adjacent to exactly two faces, and that, for any $v\in\mathscr{V}$, $\mathscr{G}\backslash v$ be connected. Geometrically, these conditions ensure that there are no loose planes or segments, and that the polyhedron is not the union of two smaller polyhedra glued together at a vertex or along an edge. Physically, such deformations would strongly cinch in the droplet surface, and are thus energetically unfavourable. Upon discarding graphs not satisfying this conditions, we are left with 44 polyhedra.

The edge graphs of these 44 polyhedra are shown in \textwholefigref{fig:polyhedra}. In particular, we find that there are only two possible polyhedra on six vertices: an octahedron, or a triple tetrahedron (that is, three tetrehadra glued together along their faces). Interestingly, random edge contractions from the icosahedron result in an octahedron with approximate probability $0.11$ only, compared to $0.89$ for the triple tetrahedron. Formation of a triple tetrahedron requires additional symmetry breaking, though. For this reason, and from this purely combinatorial analysis, the octahedron is to be expected as an intermediate step in the evolution of the icosahedron.

\section{Details of the Linear Stability Calculation}\label{appC}
In this appendix, we sketch the derivation of expressions for the different terms that appear in Eq.~(\ref{eq:Hess}). Substituting Eq.~(\ref{eq:E1b}) into expression (\ref{eq:L}) for the Lagrangian, and differentiating,
\begin{align}
\dfrac{\partial\mathcal{L}_1}{\partial\vec{x_i}}&=\vec{s_i}-\alpha\Bigl(f_\ast\vec{a_i}+f'_\ast\vec{b_i}\Bigr)-\lambda^\ast\vec{v_i},\label{eq:dL}\\
\dfrac{\partial^2\mathcal{L}_1}{\partial\vec{x_i}\partial\vec{x_j}}&=\mat{S_{ij}}-\alpha\Bigl(f_\ast\mat{A_{ij}}+f'_\ast\mat{B_{ij}}+f''_\ast\mat{C_{ij}}\Bigr)-\lambda^\ast\mat{V_{ij}},\label{eq:d2L}
\end{align}
where $f_\ast=F(\delta^\ast),f'_\ast=F'(\delta^\ast),f''_\ast=F''(\delta^\ast)$, and where the vectors $\vec{s_i},\vec{a_i},\vec{b_i},\vec{v_i}$ and matrices $\mat{S_{ij}},\mat{A_{ij}},\mat{B_{ij}},\mat{C_{ij}},\mat{V_{ij}}$ can be expressed as sums of simpler expressions, obtaining which is a mere lengthy and unpleasant exercise in differentiating vectors and their products. Solving for $\lambda^\ast$ using Eq.~(\ref{eq:dL}) and one component of $\partial\mathcal{L}_1/\partial\vec{x}=\vec{0}$ yields
\begin{align}
\lambda^\ast=s^\ast -\alpha\Bigl(a^\ast f_\ast+b^\ast f'_\ast\Bigr), 
\end{align}
where $s^\ast,a^\ast,b^\ast$ are scalars. We check \emph{post facto} that this choice of $\lambda^\ast$ indeed leads to all 36 components of $\partial\mathcal{L}_1/\partial\vec{x}$ vanishing. Hence Eq.~(\ref{eq:d2L}) becomes
\begin{align}
\dfrac{\partial^2\mathcal{L}_1}{\partial\vec{x_i}\partial\vec{x_j}}=\mat{S^\ast_{ij}}-\alpha\Bigl(f_\ast\mat{A^\ast_{ij}}+f'_\ast\mat{B^\ast_{ij}}+f''_\ast\mat{C^\ast_{ij}}\Bigr), 
\end{align}
wherein
\begin{subequations}
\begin{align}
\mat{S^\ast_{ij}}&=\mat{S_{ij}}-s^\ast\mat{V_{ij}}, &\mat{A^\ast_{ij}}&=\mat{A_{ij}}-a^\ast\mat{V_{ij}},\\
\mat{B^\ast_{ij}}&=\mat{B_{ij}}-b^\ast\mat{V_{ij}}, &\mat{C^\ast_{ij}}&=\mat{C_{ij}}.
\end{align}
\end{subequations}
Assembling these matrices into four $36\times36$ matrices made of these $3\times 3$ blocks, we obtain Eq.~(\ref{eq:Hess}).

\section{Eigenmodes of a Regular Octahedron}\label{appD}
In this appendix, we discuss the linear stability analysis of a regular octahedron. Expanding about the platonic octahedron, we may write $\mat{H}^\ddag=\mat{S^\ddag}-\alpha\bigl(F(\delta^\ddag)\mat{A^\ddag}+F'(\delta^\ddag)\mat{B^\ddag}+F''(\delta^\ddag)\mat{C^\ddag}\bigr)$, as in Eq.~(\ref{eq:Hess}), where $\delta^\ddag\approx 109.5^\circ$ is the dihedral angle of a platonic octahedron, and where the purely geometric matrices $\mat{A^\ddag},\mat{B^\ddag},\mat{C^\ddag},\mat{S^\ddag}$ are found numerically. What makes the case of the regular octahedron simpler is the fact that $\mat{A^\ddag},\mat{B^\ddag},\mat{C^\ddag},\mat{S^\ddag}$ commute pairwise, and hence can be diagonalised simultaneously~\cite{linalg}. 

\begin{figure}[b]
\includegraphics{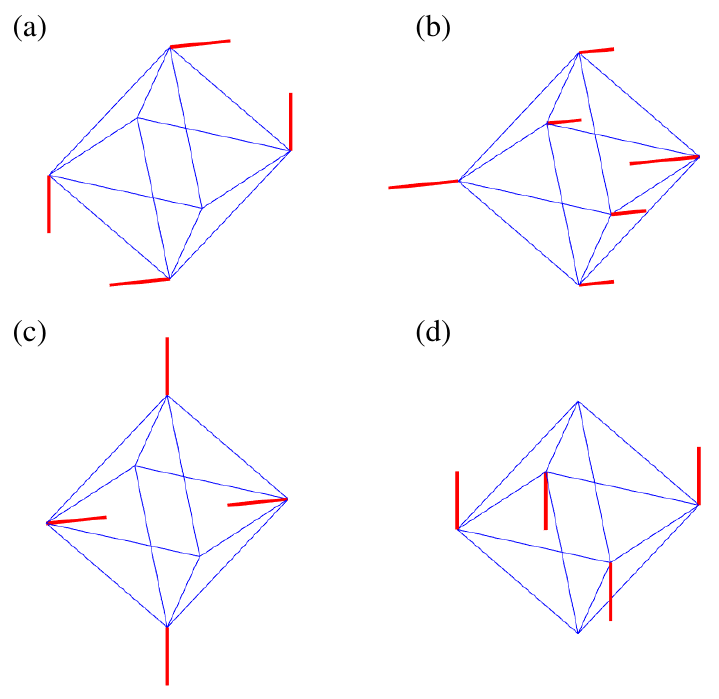}
\caption{Non-trivial eigenmodes of a regular octahedron. Dimensions of the four eigenspaces are: (a) $\dim=3$, (b) $\dim = 3$, (c) $\dim = 2$, (d)~$\dim=3$.}
\label{fig:octstab}
\end{figure}

Simultaneous diagonalisability means that stability boundaries can be computed analytically as the intersection of planes, but, here, we shall merely point out that the simultaneous eigenmodes are geometrically `nice'. Indeed, of the $6\times 3=18$ simultaneous eigenmodes~\cite{*[{Simultaneous eigenmodes and eigenvalues were computed using the algorithm described in }] [{. An implementation of the algorithm, by C. B. Mendl, was obtained from the \textsc{Matlab} file exchange (file 46794).}] simdiag}, 7 are neutral modes, corresponding to three rotations, three translations, and a scaling mode (the latter is neutral since it is not volume-preserving). The remaining 11 eigenmodes divide into four eigenspaces for which bases aligned with the symmetry axes and planes of the octahedron can be picked, as shown in \textwholefigref{fig:octstab}. 

It is natural to wonder whether there is a deeper reason for this simplification in the case of the octahedron. We do not have an answer to this question, but note that eigenmodes must respect the symmetries of the polyhedron. It is therefore tempting to speculate that, in the case of the octahedron, the existence of a common eigenbasis is caused by the fact that there are simply not enough eigenmodes that are available (i.e. allowed by the symmetries of the octahedron).

\section{Symmetric Icosahedron Model}\label{appE}
In this appendix, we derive the volume conservation constraint for the symmetric icosahedron model. Up to scaling, we may take $r^\ast=1$ for the regular icosahedron. Using an explicit coordinate representation of the icosahedron, we then obtain
\begin{align}
&R^\ast=\dfrac{\sqrt{5}-1}{2}, && H^\ast=\dfrac{\sqrt{5}+1}{4},&&h^\ast=-\dfrac{3-\sqrt{5}}{4},
\end{align}
with coordinates $r,R,H,h$ defined as in the inset of \textwholefigref{fig:icosdef}. The volume conservation constraint thus takes the form
\begin{align}
\Bigl[(r+R)^2+r^2\Bigr]H-R(2r-R)h=\dfrac{5}{2}\Bigl(\sqrt{5}-1\Bigr). 
\end{align}
We use this relation to eliminate $h$. Next, using \textsc{Mathematica} (Wolfram, Inc.), we derive expressions for the coefficients of the mobility matrix, from Eq.~(\ref{eq:M}), and for the energy gradient. These expressions, albeit too large to reproduce here, are easily evaluated numerically and given in the Supplemental Material~\cite{SM}.

%\vspace{7mm}\phantom{.}

\bibliography{drops}
\end{document}